\documentclass[10pt,prd,preprintnumbers,floatfix,aps,nofootinbib,noshowpacs,twocolumn,superscriptaddress]{revtex4} 

\usepackage{epsfig}
\usepackage{url}
\usepackage{hyperref}

\usepackage[normalem]{ulem} 

\usepackage{latexsym}
\usepackage{epsfig}
\usepackage{amsmath}
\usepackage{amssymb}
\usepackage{wasysym}
\usepackage{graphicx}
\usepackage{verbatim}
\usepackage{enumerate,mdwlist}
\usepackage[titletoc]{appendix}
\usepackage{amsfonts}
\usepackage{tikz} 
\usepackage[makeroom]{cancel}

\usepackage[normalem]{ulem}

\def\bi#1{\hbox{\boldmath{$#1$}}}

\newcommand{\FT}[1]{\mathcal{F} \{ #1\}} 

\newcommand{\zhat}{\widehat{\boldsymbol{z}}}

\newcommand{\Ex}{E(\boldsymbol{x})} 
\newcommand{\Fx}{F(\boldsymbol{x})} 

\usepackage[all]{xy} 
\usepackage{amsfonts}

\newcommand{\papertitle}{
Statistical significance testing for mixed priors: \\a combined Bayesian and frequentist analysis
}

\begin{document} 

\title{\papertitle}

\author{Jakob Robnik} \email{jakob\_robnik@berkeley.edu}
\author{Uro\v{s} Seljak}
\email{useljak@berkeley.edu}
\affiliation{Physics Department, University of California at Berkeley, \\
Berkeley, California 94720, USA}
\affiliation{Lawrence Berkeley National Laboratory, \\
Berkeley, California 94720, USA}
\begin{abstract}
In many hypothesis testing applications, we have mixed priors, with well-motivated informative priors for some parameters but not for others. The Bayesian methodology uses the Bayes factor and is helpful for the informative priors, as it incorporates Occam's razor via multiplicity or trials factor in the Look Elsewhere Effect. However, if the prior is not known completely, the frequentist hypothesis test via the false positive rate is a better approach, as it is less sensitive to the prior choice. We argue that when only partial prior information is available, it is best to combine the two methodologies by using the Bayes factor as a test statistic in the frequentist analysis. 
We show that the standard frequentist likelihood-ratio test statistic corresponds to the Bayes factor with a non-informative Jeffrey's prior. We also show that mixed priors increase the statistical power in frequentist analyses over the likelihood ratio test statistic. 
We develop an analytic formalism that does not require expensive simulations using a statistical mechanics approach to hypothesis testing in Bayesian and frequentist statistics. 
We introduce the counting of states in a continuous parameter space using the uncertainty volume as the quantum of the state. We show that both the p-value and Bayes factor can be expressed as energy versus entropy competition. 
We present analytic expressions that generalize Wilks' theorem beyond its usual regime of validity and work in a non-asymptotic regime. In specific limits, the formalism reproduces existing expressions, such as the p-value of linear models and periodograms. 
We apply the formalism to an example of exoplanet transits, where multiplicity can be more than $10^7$. We show that our analytic expressions reproduce the p-values derived from numerical simulations.  
\end{abstract}

\date{\today}

 
\maketitle

\section{Introduction}
The nature of scientific 
discovery typically proceeds 
via falsification of the null hypothesis via 
a test that is guided by an alternative
hypothesis we wish to compare to.
In many cases, we can write the alternative 
as an extension of 
the null hypothesis, such that, for example 
there is a parameter of the alternative hypothesis whose value is fixed under the null hypothesis.
In these situations, the null hypothesis is 
well specified in terms of the 
prior distributions of its parameters, while we may have little 
or no idea what the prior distribution of parameters of the alternative hypothesis is. In particular, we may have little or no idea what the prior distribution of the amplitude of the new effect should be. However, we may be performing 
the search over many parameters, some of which 
have a well-specified prior (so-called coordinate
parameters). In this paper, we are 
interested in these mixed prior situations. 

Bayesian methodology of hypothesis testing 
compares the ratio of marginal likelihoods of the two 
hypotheses to form a Bayes factor \cite{jeffreyOriginalBook}. If the prior distribution for the alternative is known, this is a valid 
methodology that yields optimal results. If
the prior
is only partially known, the resulting answer is 
sensitive to the features of the model that we 
have little control on. In particular, we can 
always arbitrarily down-weight the alternative 
hypothesis in the Bayes factor by choosing a very broad prior for 
its amplitude parameters.  
This may lead to 
rejecting the alternative when using a Bayes 
factor. This is not justifiable when the prior is poorly known: the end result can be a  
missed discovery opportunity. The opposite 
situation can also occur, where the chosen 
prior and the corresponding Bayes factor are too optimistic for the alternative 
hypothesis.

This is of most relevance when 
testing a new model (such as a new theory or a new 
phenomenon) that has 
never been observed before: 
a test 
of a new model should therefore not depend
on the reported Bayes factor alone. An alternative approach 
is that of a frequentist 
hypothesis testing, which is defined 
in terms of the false positive 
rate of the null hypothesis
(p-value, or Type I error). It is independent of the validity of the assumed
prior for the alternative hypothesis: 
we can rule out the null hypothesis 
even in the absence of a well 
developed alternative hypothesis. 
However, to do so we need
a test statistic, and the best 
test statistic is the one 
developed based on expected 
properties of alternative 
hypothesis. 

While frequentist 
false positive rate of the null 
hypothesis is a well-defined approach 
to hypothesis testing, the Bayesian 
approach has its advantages too. 
There are parameters other than 
amplitude that have 
well-specified priors, for which Bayesian hypothesis testing 
has Occam's razor built-in \cite{OccamsRazorBayesian}, and automatically 
accounts for effects such as the Look Elsewhere 
Effect \cite{LEE1}. For example, if we scan for a signal with a large template bank, we must account for 
the trials factor, 
and the corresponding p-value significance is increased \cite{Miller1981, multhyptest}. There is no single
established frequentist procedure to do this, 
while Bayes factor automatically accounts for it. 
In this case, the dependence on the prior  
in Bayes factor analysis
is an asset. In many realistic situations, the 
alternative hypothesis has several parameters, 
some with well-specified priors and some without. This paper aims 
to address these situations from both Bayesian and frequentist perspectives and propose a
solution that takes advantage of both methodologies. 

One typical example is searching for a signal in a 
time-series, such as a localized feature (e.g., planet 
transit in stellar flux): the signal could be anywhere in the time series, 
and we must pay the multiplicity penalty for looking 
at many time stamps, and Bayes 
factor automatically accounts for it. On the other 
hand, while the signal can have any amplitude,
and we may not know what amplitudes are reasonable, it is unreasonable to 
pay a high multiplicity 
penalty for a very broad amplitude prior. The Bayesian 
methodology does not have a good way to handle
this prior dependence, but frequentist methodology
does. We would like to develop a method that 
preserves the positive features of the two 
schools and removes the undesired features. 

The main themes of this paper are: 
\begin{itemize}
    \item 
 The Bayes factor is the test statistic with the highest power and should be used even in frequentist 
analyses, assuming some of the priors are 
informative (known).  
\item
For many applications, some of the priors are known, 
and some are unknown. This 
mixed prior information 
requires an analysis that combines
Bayes factor with frequentist p-value. 
\item
For mixed prior problems with some 
unknown priors frequentist p-value or Type I error (false positive rate) evaluated on the Bayes factor 
is a better 
way to summarize the significance of the alternative hypothesis than the Bayes factor itself.
In many situations, this can be done 
analytically without the simulations. 
\end{itemize}

The Neyman-Pearson lemma guarantees that
the likelihood ratio is the highest 
power test statistic for a simple alternative hypothesis with no free parameters. Similarly, the Bayes factor is the test statistic with the highest power (it minimizes the Type II error (false negative rate) at a fixed Type I error) if the alternative hypothesis has multiple parameters with some prior information. This is because the
Bayes factor summarizes optimally the prior 
information we have about the alternative 
hypothesis. It does not mean that the estimated Type II error is correct, as the prior we assumed could be wrong. It does however mean that we cannot find a better test statistic on Type II error without knowing a better prior. 

In this paper, we will offer an interpretation of the results in terms of statistical mechanics, using concepts such as 
counting of states, uncertainty quantification, and entropy. These connections have been explored
before in the context of Bayesian statistics 
\cite{jeffreypriorvolumes}, here we extend it 
to the frequentist statistics, and show that 
many of the familiar statistical mechanics
concepts can be applied to frequentist hypothesis testing. 

The main goal of this paper is to develop 
analytic methods for evaluation of p-value 
with the Bayes factor test statistic from the frequentist and Bayesian perspectives.
Our formalism enables evaluating the p-value without 
running expensive simulations, and reproduces 
many of 
the commonly used expressions in specific 
limits.\footnote{We do not discuss the 
decision theory once the p-value or the Bayes factor have been established. 
In frequentist methodology, the decision to accept 
or reject the null hypothesis is 
made with respect to some predetermined 
p-value (e.g. 0.05) to guarantee frequentist coverage, 
but choosing the 
specific value is problem and domain-dependent and its discussion is 
beyond the scope of this paper.}

The outline of the paper is as follows: 
in Section \ref{sec: bayesian testing} we define the Bayesian hypothesis testing. In Section \ref{sec: energy-entropy} we introduce the statistical mechanics approach to the Bayesian hypothesis testing. In Section \ref{sec: frequentist testing} we will develop an analytic formalism for computing the false positive rate, both from the Bayesian and frequentist perspective. In section \ref{sec: examples} we will apply this formalism to practical examples, notably an exoplanet transit search.
\section{Bayesian hypothesis testing} \label{sec: bayesian testing}

We are given some data $\bi{x}$ and want to test them against competing hypotheses. We will assume there is a single null hypothesis $\mathcal{H}_0$, which assumes there is no discovery in the data, and a collection of the alternative hypotheses $\mathcal{H}_1$, all predicting some new signal. For example, when we are looking for a planet transit in the time-series data, then the null hypothesis would be that stellar variability and noise alone are responsible for the flux variations, while the alternative hypothesis would predict an 
exoplanet transits cause some signal in the data. There are multiple alternative hypotheses because we do not know planet's properties, such as its period, phase, or amplitude, so the alternative
hypothesis is not simple, and 
has parameters we need to vary.

The Bayesian approach to hypothesis testing is to examine the ratio of the marginal likelihoods, i.e. the Bayes factor
\begin{equation} \label{eq: B def}
  B_{01}=B_{10}^{-1}
    \equiv \frac{p(\bi{x}|\mathcal{H}_0)}{p(\bi{x}|\mathcal{H}_1)},
\end{equation}
where the
marginal integral is
\begin{equation}
    p(\bi{x}|\mathcal{H}_1)=\int p(\bi{x}|\bi{z}) p(\bi{z}) d\bi{z}.
\end{equation}

Here $p(\bi{z})$ is the prior for the alternative hypothesis parameters and $p(\bi{x}|\bi{z})$ is the data likelihood under a general $\bi{z}$.
A typical situation is that the 
null hypothesis corresponds to 
some specific values of $\bi{z}$, 
such as $z_1=0$, where $z_1$ is 
the amplitude of the signal for 
the alternative hypothesis.

We are interested in the posterior odds between the competing hypothesis which follow directly from the Bayes factors by $B_{01} \, P(\mathcal{H}_0)/P(\mathcal{H}_1)$, where $P(\mathcal{H}_0)$ and $P(\mathcal{H}_1)$ are the 
prior odds. We typically assume the 
prior odds of the two 
hypotheses to be equal, in which 
case the Bayes factor is also the 
posterior odds, and in the following, we will focus on the Bayes factor only. 

The Bayes factor is an integral over all possible alternative hypotheses, but usually a relatively small range of parameters, where the data likelihood peaks, dominates the integral. In this case it suffices to apply a local integration at the peak of the posterior mass, which is often (but not always) at the maximum a posteriori (MAP) peak, where posterior
density peaks. If the location of the highest posterior mass peak under $\mathcal{H}_1$ is $\zhat$, this gives the Bayes factor as \cite{LEE1}
\begin{equation} \label{B peak}
    B_{10}
    =\frac{p(\bi{x}|\zhat)}{ p(\bi{x}|\mathcal{H}_0 )} \, p(\zhat|\mathcal{H}_1)
    V_{\rm post}(\zhat) .
\end{equation}
The Bayes factor depends on 
the likelihood ratio of the optimal model $p(\bi{x}|\zhat) / p(\bi{x}|\mathcal{H}_0 )$, prior of the MAP under $\mathcal{H}_1$ $p(\zhat|\mathcal{H}_1)$, and 
the posterior volume $V_{\rm post}(\zhat)$, which is a 
result of the local 
integration of the posterior ratio $p(\boldsymbol{z} \vert \boldsymbol{x}) / p(\zhat \vert \boldsymbol{x})$ 
around the peak. It
can be approximated using the 
Laplace approximation
as
$    V_{\rm post} \approx (2 \pi)^{d/2} \sqrt{\det \bi{\Sigma}}$,
where $\bi{\Sigma}$ is the covariance matrix, 
given by the inverse Hessian,
\begin{equation} \label{eq: Laplace covariance}
(\bi{\Sigma}^{-1} )_{i j}=-\partial_{i} \partial_{j} \ln [p(\bi{x}|\bi{z})p(\bi{z})]_{z=\widehat{z}}.
\end{equation}
Here $\partial_i$ is the derivative with respect to $z_i$. The choice
of prior is subjective, and when 
the prior is known it is informative and should be used. When the 
prior is not known, an option for a non-informative 
prior is Jeffrey's prior, which 
is given by the square root of the 
expectation of the Hessian determinant (Fisher information), and  is invariant under 
reparametrization. 
However, Jeffrey's prior also has the property
that it exactly cancels out 
the parameter dependence of $V_{\rm post}(\zhat)$ in the Bayes factor, 
making it dependent only on the 
likelihood ratio. 

We can generalize these concepts
to non-Gaussian posteriors. 
As we will show, the Laplace approximation can be poor
if the posterior is very non-Gaussian, but  
the local Bayes factor integration
is still well-defined, in which case Equation 
\eqref{B peak} can be viewed as the definition of $V_{\rm post}$ (for an example of this see Equation \eqref{eq: B Kepler}).
If the prior is $p(\bi{z}|\mathcal{H}_1) \propto 1/V_{\text{post}}(\bi{z})$ the Bayes factor is directly proportional to the likelihood ratio. Such prior choice may be called non-informative, as only the likelihood is used to test the two hypotheses. This can 
be viewed as a generalization of the Jeffrey's prior to the non-Gaussian posteriors.

\subsection{Gaussian likelihood} \label{sec: glik}

We will here make some further assumptions which simplify the analysis and will be useful in the examples we will present. Let us assume the data is a real vector $\boldsymbol{x} \in \mathbb{R}^{n}$ whose likelihood is Gaussian:
\begin{equation} \label{eq: glik}
    p(\boldsymbol{x} \vert \mathcal{H}_i) = \mathcal{N} (\bf{x} ;\, \boldsymbol{\mu}^{(i)}, \, \Sigma).
\end{equation}
The hypotheses $\mathcal{H}_0$ and $\mathcal{H}_1$ differ only in their prediction for the data mean $\bi{\mu}$. 

Since we are assuming the null hypothesis has no free parameters, we can redefine $ \bi{\Sigma}^{-1/2} (\boldsymbol{x} - \boldsymbol{\mu}^{(0)}) \xrightarrow[]{} \boldsymbol{x}$ to make the likelihood under $\mathcal{H}_0$ a standard Gaussian.

$\mathcal{H}_1$ on the other hand predicts some feature in the data:
\begin{equation}
    \boldsymbol{\mu}^{(1)} = \boldsymbol{m}(\boldsymbol{z}) = z_1 \boldsymbol{M}(\boldsymbol{z}_{> 1})
\end{equation}
where $\boldsymbol{z} = (z_1, \, z_2, \, ... \, z_d)$ is parametrizing different options in $\mathcal{H}_1$. We have split the amplitude of the signal $z_1$ from the remaining non-linear parameters $\boldsymbol{z}_{>1} = (z_2, \, ... \, z_d)$. By redefining the amplitude $z_1$, we can normalize the template: $\boldsymbol{M}(\boldsymbol{z}_{> 1}) \cdot \boldsymbol{M}(\boldsymbol{z}_{> 1}) = 1$.

We will assume that the prior variations are much slower than the likelihood variations, so the optimal model $\boldsymbol{m} (\zhat)$ is the point on the model manifold, which is the closest to the data. It has to satisfy:
\begin{equation} \label{eq: stationarity}
    (\boldsymbol{x} - \boldsymbol{m}) \cdot \partial_i \boldsymbol{m} = 0 \qquad i = 1, \, 2, \, ... \, d .
\end{equation}

The log-likelihood ratio is related to the $\chi^2$ improvement of the fit:
\begin{align} \label{eq: simplified chi2}
    E &\equiv \log \frac{p(\bi{x}|\zhat)}{ p(\bi{x}|\mathcal{H}_0 )} = \frac{1}{2} \widehat{z}_1^2,
\end{align}
where we have used that $\boldsymbol{x} \cdot \boldsymbol{m} (\zhat) = \boldsymbol{m}^2 (\zhat)$ by Equation \eqref{eq: stationarity}. We can identify $\widehat{z}_1$ as the signal-to-noise-ratio. 

The posterior is in general not Gaussian and the Laplace approximation may be a poor 
approximation (e.g. Section \ref{sec: exoplanet}). Here we adopt it anyways to gain some intuition. Equation \eqref{eq: Laplace covariance} gives the posterior covariance matrix:
\begin{equation} \label{eq: Fisher covariance}
    \Sigma^{-1}_{i j} = \partial_i \boldsymbol{m} \cdot \partial_j \boldsymbol{m} - (\boldsymbol{x} - \boldsymbol{m}) \cdot \partial_i \partial_j \boldsymbol{m}.
\end{equation}
The first term is independent 
of the data and thus equals the Fisher information matrix (which 
is defined as an average over 
data realizations)
\begin{equation} \label{eq: Fisher def}
    I_{ij} = \partial_i \boldsymbol{m} \cdot \partial_j \boldsymbol{m},
\end{equation}
while we drop the second term since $\boldsymbol{x} - \boldsymbol{m}$ is rapidly oscillating around zero if there are no systematic residuals and the Hessian of the model is typically varying more slowly (Gauss-Newton 
approximation). 
Although the Bayes factor is in general a function of the data $\boldsymbol{x}$, we have here approximated it solely as a function of the best fit parameters $\zhat$, inferred from the data:
\begin{equation} \label{eq: B glik}
    B_{10}(\zhat) = \frac{(2 \pi)^{d/2} p(\zhat | \mathcal{H}_1)}{[ \det I(\zhat) ]^{1/2} } \exp \bigg{\{} \frac{1}{2} \widehat{z_1}^2 \bigg{\}}.
\end{equation}
Note that the Fisher determinant can be restricted to the $\boldsymbol{z}_{>1}$ components only, since there are no correlations between $z_1$ and the other parameters: $I_{1 i} =  \boldsymbol{M} \cdot z_1 \partial_i \boldsymbol{M} = z_1 \partial_i \boldsymbol{M}^2 = 0$ and $I_{1 1} = \boldsymbol{M} ^2 = 1$. This property is exact and does not rely on the Laplace approximation, as can be seen by using the Gaussian likelihood \eqref{eq: glik} in the Bayes factor computation \eqref{eq: B def}.
We also note that the errors on the non-linear parameters scale as the inverse signal-to-noise-ratio, since $I_{i j} = z_1^2 \, \partial_i \boldsymbol{M} \cdot \partial_j \boldsymbol{M}$, but this is only true under the Fisher approximation to the posterior.

Many cases of practical interest fall under the assumptions in the present section, often with some necessary preprocessing. For example, the Kepler space telescope data are not Gaussian distributed because of the outliers, but they can be Gaussianized \cite{GMF}. We will show multiple applications of this setup throughout the paper: linear models in Subsection \ref{sec: linear 1} and \ref{sec: linear 2}, periodogram in Subsection \ref{sec: periodogram 1} and \ref{sec: periodogram 2} and exoplanet transit search in Section \ref{sec: exoplanet}.

\section{Energy-Entropy decomposition} \label{sec: energy-entropy}

In a continuous parameter space, the nearby models are not independent. We can think of two models as indistinguishable if we cannot favor one or the other after seeing the posterior. This suggests that all models within a posterior volume can be considered one unit, which we call a state\footnote{Equivalently one can define a state as a collection of models whose relative entropy is small, see \cite{jeffreypriorvolumes}.}. This is analogous to the shift from classical statistical mechanics to quantum statistical mechanics, where the discrete states are counted in units of their uncertainty volume. 

Let's first consider the Jeffrey's prior choice, which is non-informative in the sense that it assigns an equal probability to each state \cite{jeffreypriorvolumes}, so to each indistinguishable model. The logarithm of the Bayes factor 
\begin{equation}
    F \equiv \log B_{10} = E - \log \frac{1}{p(\zhat| \mathcal{H}_1) V_{\rm post}(\zhat)} = E - \log N \equiv E - S
\end{equation}
is reminiscent of the thermodynamics relation for the free energy if we identify $E$ with energy, since the entropy $S = \log N$ is independent of $\zhat$ and equals the logarithm of the number of states in the prior volume. For a more general prior, the thermodynamics relation has to be generalized to $F = E + U - S$, where the potential energy $e^{U} =  \frac{p(\boldsymbol{z})}{p_{Jeff}(\boldsymbol{z})}$ is given by the prior relative to the Jeffrey's prior and the total energy is $E + U$.

Our definition of energy and entropy should be viewed from the hypothesis testing point of view, where the energy $E + U$ is the only "macroscopic" parameter that influences the outcome of the test. The other parameters are "microscopic" in the sense that we do not care about their values in the test. Entropy is the logarithm of the number of microstates, given the macro state, as usual. To be precise, the entropy should only count the states which give rise to the same macro state, so the states with the same energy. Such count corresponds to the Bayes factor which ignores the look-elsewhere effect associated with the amplitude parameter; we will call it $B_{>1}$. It is not justified from the Bayesian perspective as the prior would have to be fixed after seeing the data. Still, it makes more sense from the statistical mechanics point of view, and we will return to this quantity when it appears in the frequentist analysis in the next section.

Note that the entropy is always positive and therefore always favors $\mathcal{H}_0$, since the posterior is narrower than the prior. Energy has to surpass entropy for the alternative hypothesis to prevail. This is the Occam's razor penalty, which is built into the Bayes factor. 
The main deficiency of the Bayes factor is in situations where the prior of the alternative hypothesis is poorly known, as is invariably the 
case when we encounter a new phenomenon we have 
never seen before. In this case  
Bayes factor is sensitive to the assumption on the prior of 
alternative hypothesis, 
which can be arbitrary, as we will now illustrate with two examples.

\subsection{Example: linear model} \label{sec: linear 1}

First consider the setup in Subsection \ref{sec: glik} and further assume $\mathcal{H}_1$ models the data as a linear superposition of $d$ features $\boldsymbol{m}_i \in \mathbb{R}^n$:
\begin{equation} \label{eq: m linear}
    \boldsymbol{m}_{\text{linear}}(\boldsymbol{w}) = \sum_{i = 1}^d w_i \boldsymbol{m}_i
\end{equation}
Without loss of generality, we may assume the features are orthonormal $\boldsymbol{m}_i \cdot \boldsymbol{m}_j = \delta_{i j}$ by applying the Gram-Schmidt algorithm if they are not.

The optimal model is  a projection of the data on the model plane $\widehat{w}_i = \boldsymbol{x} \cdot \boldsymbol{m}_i$ and has the log-likelihood ratio $E = \frac{1}{2}\sum_{i =1}^d \widehat{w}_i^2$. The posterior is Gaussian, so the Laplace approximation is exact and gives the posterior volume $(2 \pi)^{d/2}$.

Often we do not have any prior information, and it makes sense to adopt the Jeffrey's prior, which is uniform for the linear parameters. For simplicity, we assume the prior volume to be a ball with $\sum_{i=1}^d w_i^2 < w_{max}^2$, but we do not know how to choose $w_{max}$. We cannot just send it to infinity since the Bayes factor
\begin{equation} \label{eq: B linear}
    B_{10} = \frac{(2 \pi)^{d/2} e^E}{V(B^d) w_{max}^d }
\end{equation}
would go to zero. Here, $V(B^d)$ is the volume of the $d$ dimensional unit ball. We will show another, more realistic, example with the same problem.

\subsection{Example: periodogram} \label{sec: periodogram 1}
In this example, we are given $n$ time series measurements $x_i = x(t_i)$ with uncertainties $\sigma$. We are 
searching for harmonic periodic signals \cite{Lomb1976, Scargle1982,VanderPlas2018}: 
\begin{equation} \label{eq: m periodogram}
    m_i(\bi{z})= A \sin(\omega t_i + \phi)
\end{equation}
Here $A$ is the signal's amplitude, $\omega$ is the unknown frequency, and $\phi$ is the phase.

To make the calculations tractable we assume dense data sampling over many periods of the signal, so that we can replace the discrete sum in the scalar products of Subsection \ref{sec: glik} with the integral $\sum_{i = 1}^n \xrightarrow[]{} \frac{n}{T} \int_0^T dt$, where $T$ is the total time of the measurements. We will work with the amplitude $z_1$ relative to the normalized template, which has $A  = \sigma \sqrt{2 / n}$.
We assume a uniform prior on all three parameters, so 
\begin{equation}
    p(z_1, \omega, \phi) = \frac{1}{z_{max} \, \omega_{max} \, 2 \pi},
\end{equation}
where $z_{max}$ is some arbitrary large cutoff on the amplitude and $\omega_{\rm max}$ is set by some physical properties of the signal or by the experimental limitations, e.g. by the Nyquist frequency.

The inverse of the Fisher information matrix \eqref{eq: Fisher def} is
\begin{equation} \label{eq: I periodogram}
    I^{-1} = 
    \begin{bmatrix}
    \sigma_{\omega}^2 & \sigma_{\omega \phi} \\
    \sigma_{\omega \phi} & \sigma_{\phi}^2
    \end{bmatrix} = 
    E^{-1} \begin{bmatrix}
    6 / T^2 & - 3/ T \\
    -3 / T & 2
    \end{bmatrix} .
\end{equation}
Interestingly, the frequency-phase correlation is relatively high $\sigma_{\omega \phi} / \sigma_{\omega} \sigma_{\phi} = -\sqrt{3} / 2 \approx -0.87$. The Bayes factor
\begin{equation} \label{eq: B periodogram}
    B_{10} = \frac{\sqrt{6 \pi} \,  e^{E}}{z_{max} \, \omega_{max} T\, E},
\end{equation}
again suffers from the unknown amplitude cutoff. If we set it too high, we might reject a true periodic
signal.

To overcome the arbitrariness of the prior choice, one could try to learn the prior from the data under some hyper-parametrization. This leads to a hierarchical or empirical Bayesian analysis \cite{hierarchicalBayes, empiricalBayes}, but this is not possible if we have no data that can inform the hyperparameters.

Note that the constant energy Bayes factor $B_{>1}$ is independent of the unknown cutoff in both examples, but we have not yet justified its use. We will see its appearance when we phrase the test in terms of the frequentist false positive rate, which we discuss now.

\section{Frequentist hypothesis testing} \label{sec: frequentist testing}

In frequentist hypothesis testing, we first define 
a test statistic, which should 
be chosen to maximize the contrast against the 
alternative hypothesis. As argued in the previous 
section, the optimal statistic between two well-specified hypotheses are the posterior odds. Any monotonically increasing function of the posterior odds is an equally good test statistic in the frequentist sense, so for convenience, we will work with $F = \log B_{10}$.
If we adopt the generalized Jeffrey's prior, then $p(\bi{z}|\mathcal{H}_1) V_{\text{post}}(\bi{z})$ is a constant and 
the test statistic becomes the likelihood ratio. This is the most common frequentist test statistics, but 
it is not prior independent. Instead, it corresponds to a specific prior choice - the generalized Jeffrey's prior.
Prior information about some parameters can reduce the Type II error at a constant Type I
error by the Neyman-Pearson argument, so in general, we will work with $F$ as the optimal test statistic.  

In frequentist methodology, 
we quantify a test statistic using
its Type I false positive rate (p-value) under the null hypothesis $\mathcal{H}_0$. 
While $\mathcal{H}_1$ is used to define 
the test statistic, the distribution of the test 
statistics under $\mathcal{H}_0$ is independent of whether 
$\mathcal{H}_1$ distribution is true or not.
If this number is sufficiently 
small, the null hypothesis is rejected. Unlike the posterior odds, the p-value is independent of the correctness of the prior of the alternative hypothesis, such as the 
value of $w_{\rm max}$: it only depends on the null 
hypothesis itself. 
This is a great advantage
of p-value over the posterior odds, and is the main 
reason why posterior odds have not been widely adopted for hypothesis testing even in Bayesian textbooks \cite{BayesFactorOverview}.
 
While the posterior odds can 
be derived entirely from the data using the 
likelihood (the  likelihood principle), the p-value 
is given by the frequency distribution of the test statistic
under the null hypothesis, which generally requires simulating the null hypothesis many times to obtain it. Since this requires evaluating the test statistic for each simulation, and since we argue 
the test statistic itself should be the posterior odds, this 
can be significantly more expensive in 
frequentist methodology 
than evaluating posterior odds once on the data, as done in
Bayesian methodology. For this reason, we will study analytic techniques for evaluating the false-positive rate.

Given the choice of the test statistic $F$, the 
frequentist analysis proceeds to quantify the frequency of null events above its observed value $\Fx$ in a frequentist sense, which means under a long series of trials. 
This gives the p-value as a 
function of the observed test statistic 
$\Fx$, which quantifies
the false positive rate (FPR) 
as $P(F > \Fx)$.  
 
 \subsection{p-value from frequentist property of Bayes factor}
We will assume that the Bayes factor can be expressed as a function of the optimal inferred parameters $\zhat$, as in Equation \eqref{eq: B glik}. Then the distribution of $F$ can be inferred from the distribution of the optimal parameters. For the 
null hypothesis, the majority of events will 
have optimal parameters close to the null hypothesis values, 
but there will be outliers whose frequency we would 
like to evaluate. If the prior is correct, the Bayes factor predicts the relative frequency $\mathcal{P}$ of the outcomes
of the two hypotheses in a 
long series of trials:
\begin{equation} \label{eq: null events}
    \mathcal{P}(\zhat \vert \mathcal{H}_0) = \mathcal{P}(\zhat \vert \mathcal{H}_1) B_{0 1} (\zhat) \approx  p(\zhat |\mathcal{H}_1) e ^ {- F(\zhat)}
\end{equation}
The frequency of $\mathcal{H}_1$ events follows the prior: each sample with inferred MAP parameters $\zhat$ will have a true value of parameters $z$ approximately within $V_{\rm post}$ of $\zhat$. Thus, as long as the prior is sufficiently smooth relative to the posterior, we have $\mathcal{P}(\zhat \vert \mathcal{H}_1) \approx p(\zhat | \mathcal{H}_1)$.

The p-value is found by integrating over the parameters which yield the desired test statistic:
\begin{align} \label{eq: main result}
    P(F >  \Fx) &= \int_{F(\zhat) > \Fx} p (\zhat| \mathcal{H}_1) e^{-F(\zhat)}  d\zhat = \\
    &= \int_{\Fx}^{\infty} p(F) e^{-F} dF,
\end{align}
where $p(F)$ is the marginal prior density. This result is saying that the probability density of finding some $F$ under the null hypothesis is given by the probability density of finding it under the alternative hypothesis prior, but exponentially damped with $F$.
Note that Equation \eqref{eq: main result} is equivalent to
\begin{equation} \label{eq: prior dependence}
    P(F >  \Fx) = \int_{F(\zhat) > \Fx} \frac{e^{-E(\zhat)}}{V_{\text{post}}(\zhat)} d\zhat ,
\end{equation}
so the region of integration depends on the prior, but the integrand does not. This means that the prior selects the parameter range where the false positives can be generated, but the rate at which they are generated at those parameters is prior-independent.

These expressions are useful when we have an analytic expression for the posterior volume and can perform the integral analytically. We will show examples of this situation in Section \ref{sec: examples}. 
Here we will derive a useful approximation which relates the p-value and the constant energy Bayes factor $B_{>1} = e^{E} p_{>1}(\widehat{\boldsymbol{z}}) V_{\text{post}}^{(>1)}(\zhat)$. This will enable the p-value calculation even when the Bayes factor is only available numerically. 
Note that the variations of $V_{\text{post}}$ as a function of $E$ are much slower than the suppression $e^{-E}$. One can therefore consider the posterior volume to be a constant evaluated at $z_1 = \sqrt{2 E(\boldsymbol{x})}$ when taking the $z_1$ integral. This gives
\begin{equation} \label{eq: approx pvalue}
    P(F > \Fx) \approx \frac{e^{-\Ex}}{\sqrt{4 \pi \Ex} \, p_{>1}(\widehat{\boldsymbol{z}}) V_{\text{post}}^{(>1)}(\zhat)} = \frac{1}{B_{>1} \sqrt{4 \pi E}}
\end{equation}
which is approximately valid for small p-values. This expression implies that the p-value can be directly inferred from the constant energy Bayes factor $B_{>1}$ with no need to do any additional integrals.

\subsection{Frequentist derivation of p-value}

\begin{figure}
    \centering
    \includegraphics[scale = 0.28]{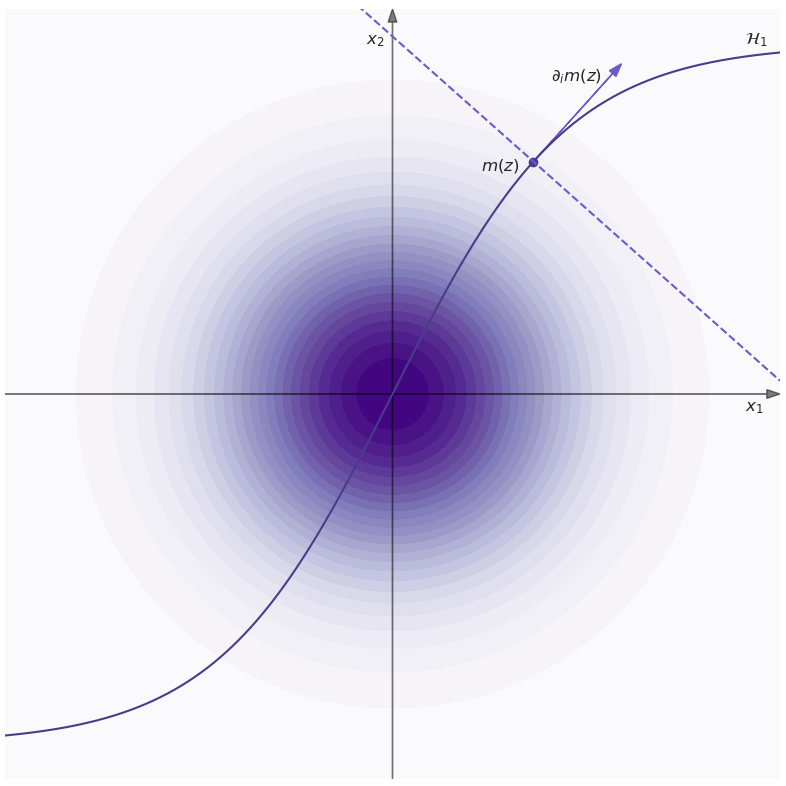}
    \caption{Illustration of a two-dimensional data space ($n = 2$). The concentric circles are the likelihood contours of $\mathcal{H}_0$. The solid line is the $\mathcal{H}_1$ manifold ($d = 1$). If the data point $\boldsymbol{x}$ lies on the dashed line, the model $\boldsymbol{m}(\boldsymbol{z})$ satisfies Equation \eqref{eq: stationarity} and is at least locally the optimal model. Integral of Equation \eqref{eq: plane integral} runs over the dashed line.}
    \label{fig: model manifold}
\end{figure}

We will here present an approach that is complementary to the previous section in the sense that it does not use any of the Bayesian concepts but still reproduces the same results. Since there is no concept of a prior, we will be working with the likelihood ratio as a test statistic (which is the Bayes factor with the Jeffrey's prior). We will assume the setup as in Subsection \ref{sec: glik}. 

First, let us determine how likely would we have seen some model $\boldsymbol{m}(\boldsymbol{z})$ as the optimal model if the null hypothesis was true. We add the probabilities over the suitable region of the data space:
\begin{equation} \label{eq: N integral}
    p(\boldsymbol{m}(\boldsymbol{z}) \vert \mathcal{H}_0) = \int_{\mathcal{N}} p(\boldsymbol{x} \vert \mathcal{H}_0) d\boldsymbol{x},
\end{equation}
where $\mathcal{N}$ is the set of all points which have $\boldsymbol{m}(\boldsymbol{z})$ as their optimal model. $\mathcal{N}$ is a subset of the plane defined by Equation \eqref{eq: stationarity}.This equation defines an affine $n -d$ dimensional plane which is an orthogonal complement to the tangent plane to the $\mathcal{H}_1$ manifold at the point $\boldsymbol{m}(\boldsymbol{z})$ (for illustration, see Figure \ref{fig: model manifold}). However, $\mathcal{N}$ is not the entire plane since Equation \eqref{eq: stationarity} is a necessary but not sufficient condition. In other words, the posterior may be multi-modal. To make progress, we will have to neglect this and assume $\mathcal{N}$ to be the entire plane. By doing so, our final result for the p-value is in fact an upper bound, which becomes more and more accurate the lower the p-value.

The model $\boldsymbol{m}(\boldsymbol{z})$ is both a point on the plane $\mathcal{N}$ and a normal to the plane, as can be seen from Equation \eqref{eq: stationarity} with $i = 1$. Writing $\boldsymbol{x} = \boldsymbol{m}(\boldsymbol{z}) + \boldsymbol{x}_{\bot}$ for points on the plane and using orthogonality, we can easily evaluate the Gaussian integrals \eqref{eq: N integral}:
\begin{equation} \label{eq: plane integral}
    p(\boldsymbol{m}(\boldsymbol{z}) \vert \mathcal{H}_0) = \int p(\boldsymbol{x} \vert \mathcal{H}_0) d\boldsymbol{x}_{\bot} = \frac{e^{-\boldsymbol{m}^2 / 2}}{(2 \pi)^{d /2}}.
\end{equation}
This result is saying that the probability density in the data space of a model occurring as an optimal model under $\mathcal{H}_0$ depends only on $\boldsymbol{m}^2 / 2 = E$. Therefore, the probability of observing $E$ under the null hypothesis equals the probability of finding any $\boldsymbol{m}$ with $\boldsymbol{m}^2 / 2 = E$ as the optimal model. This gives
\begin{equation} \label{eq: frequentist fpr}
    p(E \vert \mathcal{H}_0) = \frac{1}{\sqrt{2 E}}\frac{e^{ - E}}{(2 \pi)^{d /2}} V_{\text{shell}},
\end{equation}
where $V_{\text{shell}}$ is the data space volume of models with $\boldsymbol{m}^2 / 2 = E$, in other words, the constant likelihood surface volume. We have picked up a factor $(2 E)^{-1/2}$ when transforming the density with respect to $\widehat{z_1}$ to the density with respect to $E$. 
The volume of the constant likelihood shell in the data space is an integral over the $\mathcal{H}_1$ manifold at fixed $z_1 = \sqrt{2 E}$. It can be evaluated in the $\boldsymbol{z}_{>1}$ parameter space by integrating the square root of the determinant of the metric over the coordinate range:
\begin{equation} \label{eq: manifold volume}
    V_{\text{shell}} = \int \, d \boldsymbol{m}(\sqrt{2 E}, \boldsymbol{z}_{>1}) = \int \det I_{i j} (\sqrt{2 E}, \boldsymbol{z}_{>1})^{1/2} d \boldsymbol{z}_{>1},
\end{equation}
where the metric on the $\mathcal{H}_1$ manifold coincides with the Fisher information matrix \eqref{eq: Fisher def}. Combining Equations \eqref{eq: frequentist fpr} and \eqref{eq: manifold volume} we find
\begin{equation} \label{eq: E pdf glik}
    p(E \vert \mathcal{H}_0) = \frac{e^{ - E}}{\sqrt{2 E}} \int \frac{d \boldsymbol{z}_{>1}}{V_{\text{post}} (\sqrt{2 E}, \boldsymbol{z}_{>1}) },
\end{equation}
where $V_{\text{post}} = (2 \pi)^{d/2} \det I_{i j}^{-1/2}$.

Even though we use frequentist formalism, the posterior volume enters as the determinant of the transformation from the data space to the parameter space. We reproduce the result of Equation \eqref{eq: prior dependence}.

There are some exciting applications where the derived expressions are exact (see Section \ref{sec: linear 2}) since the posterior has a single mode. However, in general, the frequentist derivation made it manifest that in the presence of multiple modes, the p-value results are only an upper bound, which becomes useless if the p-value is high. We now extend our results to all p-values.

\subsection{Non-asymptotic p-value} \label{sec: sidak}
We have found an analytic expression for the false positive rate, both from the frequentist and Bayesian perspectives. However, it is only valid when the false positive rate is low. We will show that this result can be easily extended to all p-values if the posterior volume is much smaller than the prior volume, as was done in \cite{LEE1}.

Let's partition the alternative hypothesis manifold in $K$ smaller manifolds and consider searches over the smaller manifolds. Let's choose the partition in a way that FPR in all the small searches is the same and call it $P_{\text{small}} (F > \Fx)$. 

Suppose the posterior volume is much smaller than the prior volume. In that case, most data realizations will have their posterior volume well within the prior range of some smaller manifold, even when $K$ is relatively large. Therefore, the probability of not finding a false positive in the original search equals the probability of not finding it in any of the small searches: 
\begin{equation}
    1 - P(F > \Fx) = \big{\{} 1 - P_{\text{small}}(F > \Fx) \big{\}} ^K,
\end{equation}

If $K$ is large, $P_{\text{small}}(F > \Fx)$ becomes small, and we can compute it using the asymptotic expression of Equation \eqref{eq: main result} or \eqref{eq: approx pvalue}: $P_{\text{small}} (F > \Fx) = P_{\text{asym}} (F > \Fx) / K$. Taking the large $K$ limit we find
\begin{equation} \label{eq: sidak}
    P(F > \Fx) = 1 - \exp \{- P_{\text{asym}}(F > \Fx) \},
\end{equation}
This is a continuous parameters generalization of Sid\' ak correction, which itself
is a generalization of Bonferroni correction, which 
are commonly used for discrete states where the trials factor 
is a well-defined concept and referred to as 
multiple test comparison. 

\subsection{Statistical mechanics interpretation} \label{sec: stat mech}
\begin{figure*}
    \begin{minipage}{0.4\textwidth}
    \includegraphics[scale = 0.29]{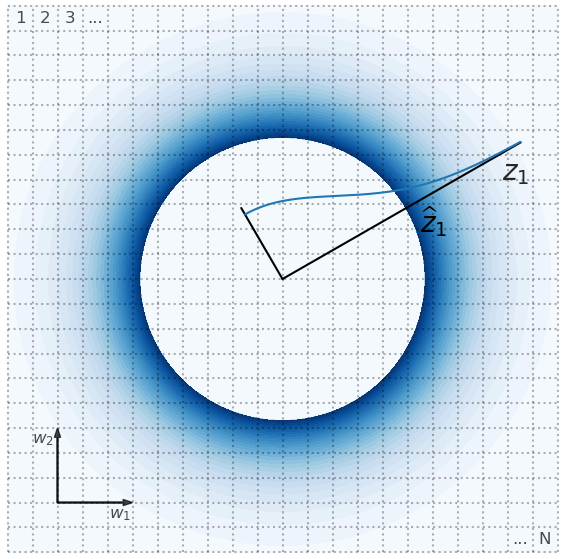}
    \end{minipage}
    \begin{minipage}{0.4\textwidth}
    \includegraphics[scale = 0.33]{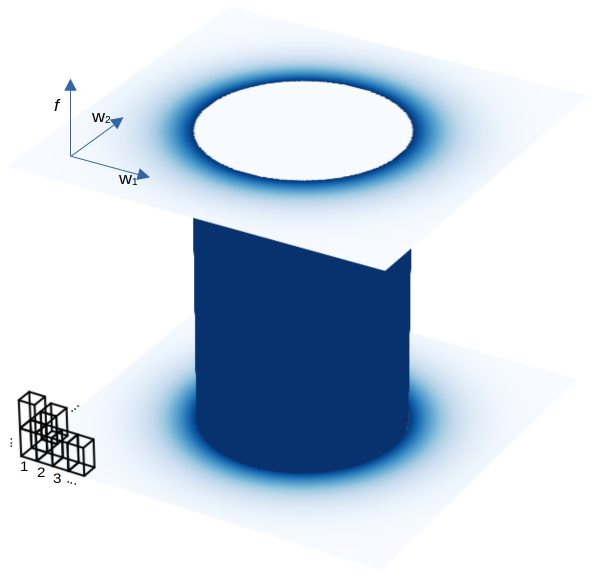}
    \end{minipage}
    \caption{Illustration of the statistical mechanics interpretation of the p-value. We emphasize the difference between the amplitude and the other parameters. 
    Left panel: parameter space of a linear model \eqref{eq: m linear} with two degrees of freedom. This is equivalent to the periodogram model \eqref{eq: m periodogram} at a fixed frequency $\omega$. Constant energy (likelihood) shells are circles of radius $z_1 = (w_1^2+w_2^2)^{1/2}$. We assume a uniform prior with equal uncertainty on both variables and show posterior volume as the area of each mesh cell. The expected ratio of the null hypothesis to the alternative hypothesis events in each cell is given by the Boltzmann factor in Equation \eqref{eq: partition function}. It dies off exponentially as a function of expected energy. The color intensity in the plot is proportional to this Boltzmann factor, which is shown along one radial direction as a blue line. False positive rate counts all the states in the region exterior to the circle of observed $\Fx$ and weights them with the Boltzmann factor \eqref{eq: partition function}. Only the region close to the observed test statistic circle will contribute due to the exponential suppression. Increasing the prior on $w_1$ and $w_2$ increases the total number of states and reduces the Bayes factor but has no impact on the p-value \eqref{eq: p linear}.
    Right panel: now, we also vary the frequency, searching for sinusoidal signal \eqref{eq: m periodogram} over all frequencies within some prior range. The observed $\Fx$ surface is a cylinder and the false positive rate is proportional to its surface area. Increasing the prior range of $\omega$ over which we search for sinusoidal signal reduces the Bayes factor because additional states were introduced \eqref{eq: B linear}. Contrary to the first example, the false positive rate has now also increased \eqref{eq: p periodogram} because some new states are close to the observed $\Fx$ shell.
} \label{fig: multiplicity}
\end{figure*}
Returning to the discrete picture of the continuous parameter space presented in Section \ref{sec: energy-entropy} we recognize the integral $\int \frac{d \boldsymbol{z} }{V_{\text{post}}(\boldsymbol{z})}$ in Equation \eqref{eq: prior dependence} as the the continuous version of the sum over states. The p-value is therefore the sum over all states which generate false positives, each weighted with the Boltzmann factor $e^{-E}$. The p-value is therefore the partition function of the canonical ensemble with $k_B T = 1$:
\begin{equation} \label{eq: partition function}
    P(F >  \Fx) = \sum_{F(\text{state}) > \Fx} e^{-E(\text{state})} = e^{-F_{f}},
\end{equation}
where $F_{f}$ is the frequentist free energy. It is evident why the p-value is not sensitive to the high amplitude cutoff: high energy states are suppressed by the Boltzmann factor and do not contribute to the p-value. An illustration of this interpretation is presented in Figure \ref{fig: multiplicity}.

Note that the p-value approximation \eqref{eq: approx pvalue} implies that the Bayesian free energy $\log B_{>1}$ and the frequentist free energy $- \log P(F > \Fx)$ are one and the same thing, up to logarithmic corrections in energy.
Note that in physics, the thermodynamic and statistical mechanic free energies are also the same in the thermodynamic limit.

\section{Results} \label{sec: examples}

\begin{figure*}
    \centering
    \hspace*{-0.3cm}\includegraphics[scale = 0.13]{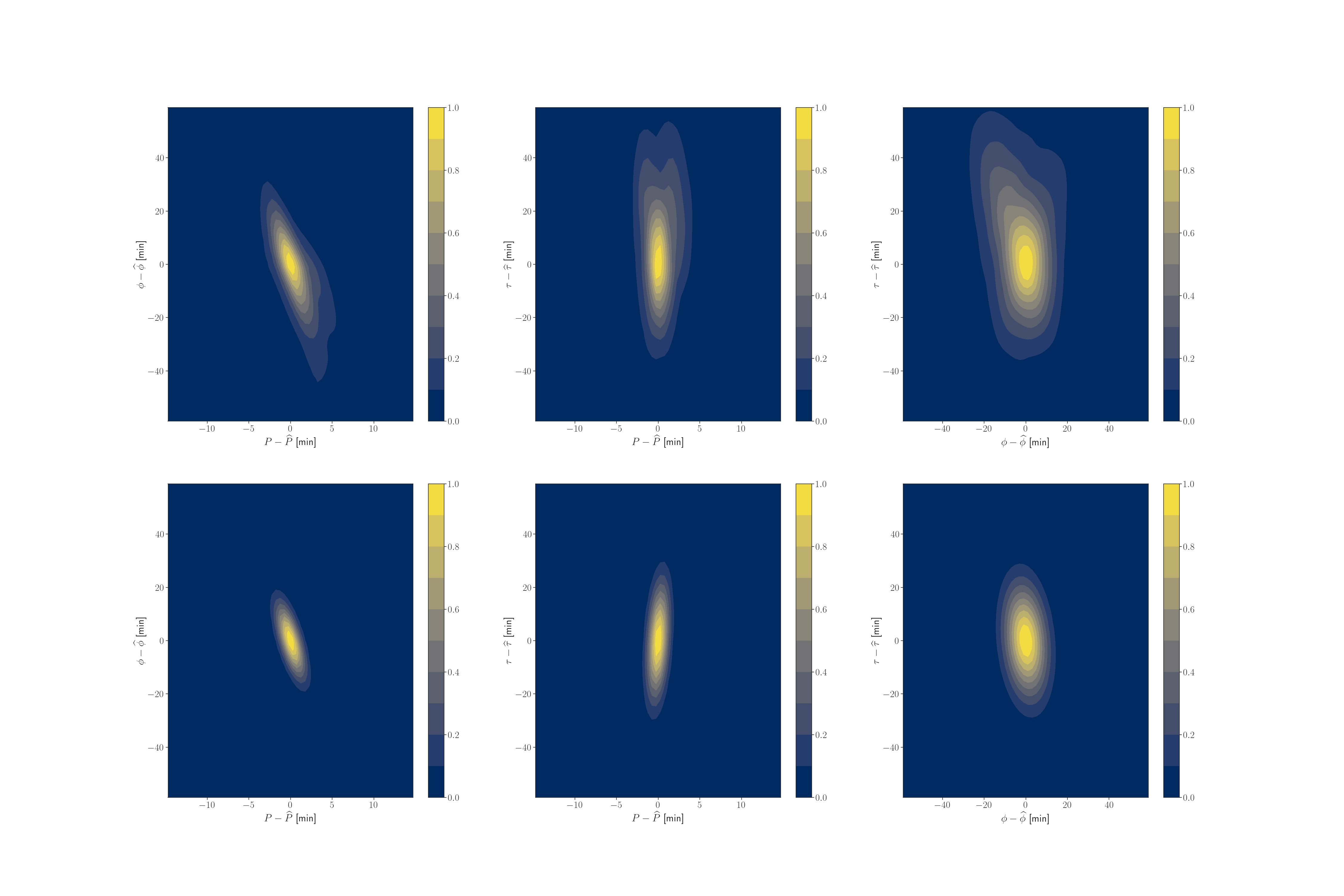}
    \caption{We show an example of a typical noise simulation with the injected planet with a period $P = 100$ days and transit duration $\tau = 0.4$ days. We show the likelihood ratio in the neighborhood of the MAP parameters $\zhat$ (upper panels). Laplace approximation expands the likelihood ratio's logarithm to a quadratic order and approximates the peak with a Gaussian (lower panels). Note that the Laplace approximation is poor, corresponding to an inaccurate p-value estimation (Figure \ref{fig: pvalues}).}
    \label{fig: integrand}
\end{figure*}

In this section, we apply the developed formalism to three examples with increasing complexity. We start with the linear model and periodogram and compare our results with the literature. We then turn to the more realistic analysis of the exoplanet transit search.

\subsection{Linear model} \label{sec: linear 2}
We start with the linear model as in Section \ref{sec: energy-entropy}. As seen from Equation \eqref{eq: B linear} $F$ and $E$ only differ by a constant, so they are equally good test statistics, and we might as well work with $E$. The posterior volume is a constant, so the integral of Equation \eqref{eq: prior dependence} just picks up the volume of the constant energy surface, which is a $d -1 $ dimensional sphere. We get
\begin{equation} \label{eq: p linear}
    P(E > \Ex) = \int_{\Ex}^{\infty} \frac{E^{d/2 -1} \, e^{-E}}{ \Gamma (d/2)} dE,
\end{equation}
where we have used that the volume of a $d-1$ dimensional sphere is $2\pi^{d/2}/\Gamma(d/2)$, with $\Gamma$ the Gamma function. 

The resulting cumulative distribution function is a $\chi^2$-distribution with $d$ degrees of freedom, and we reproduce the well-known result that $\chi^2$ of a linear model with $d$ features is distributed as a $\chi^2$ distribution with $d$
degrees of freedom \cite{wilksTheorem}. The p-value is
increasing with $d$ at a constant $ \chi^2=2 E$, which is a reflection 
of the entropy versus energy competition: there  
are more states on the shell of a constant energy $E$
if $d$ is higher. 

Note that the p-value is independent of the unknown amplitude cutoff $w_{max}$.


\subsection{Periodogram} \label{sec: periodogram 2}
We now turn to the periodogram example, which is defined in Section \ref{sec: periodogram 1}. We have seen in Equation \eqref{eq: B periodogram} that the Bayes factor and the likelihood ratio are a simple function of one another, so their null distributions are related by the change of variable formula $p(F \vert \mathcal{H}_0) = \vert d F / d E \vert \,  p(E \vert \mathcal{H}_0) = \vert 1 - 1/E \vert \,  p(E \vert \mathcal{H}_0)$. 

Once again, the posterior volume is independent of the $\boldsymbol{z}_{>1}$ parameters, and the integral \eqref{eq: prior dependence} picks up the volume of the constant energy surface, which is, in this case, a cylinder, so $2 \pi \omega_{max}$. We get
\begin{equation} \label{eq: p periodogram}
p(E \vert \mathcal{H}_0) = \frac{1}{B_{>1}\,  \sqrt{4 \pi E}}  ,
\end{equation}
so the false positive rate is closely related to the Bayes factor, as in the approximation \eqref{eq: approx pvalue}. Note that while the amplitude cutoff cancels in the p-value, the frequency cutoff does not. This is the look-elsewhere effect: the larger the frequency space search, the larger the false positive rate at a fixed likelihood-ratio.

These results agree with expressions of \cite{Baluev_2008}, which use the formalism of \cite{Davies,Davies2}. This mathematical formalism is based on extremes of random processes of various distributions 
such as gamma ($\chi^2$) distribution, and needs to be derived
separately for each distribution. In 
addition, this formalism formally only gives an analytic lower limit to the corresponding extreme value distributions, while in practice, 
equality is assumed
without proper justification. Our 
formalism provides a different derivation 
that results in the same expressions in the periodogram case.

\subsection{Exoplanet search} \label{sec: exoplanet}

\begin{figure*}
    \centering
    \hspace*{-1cm}\includegraphics[scale = 0.4]{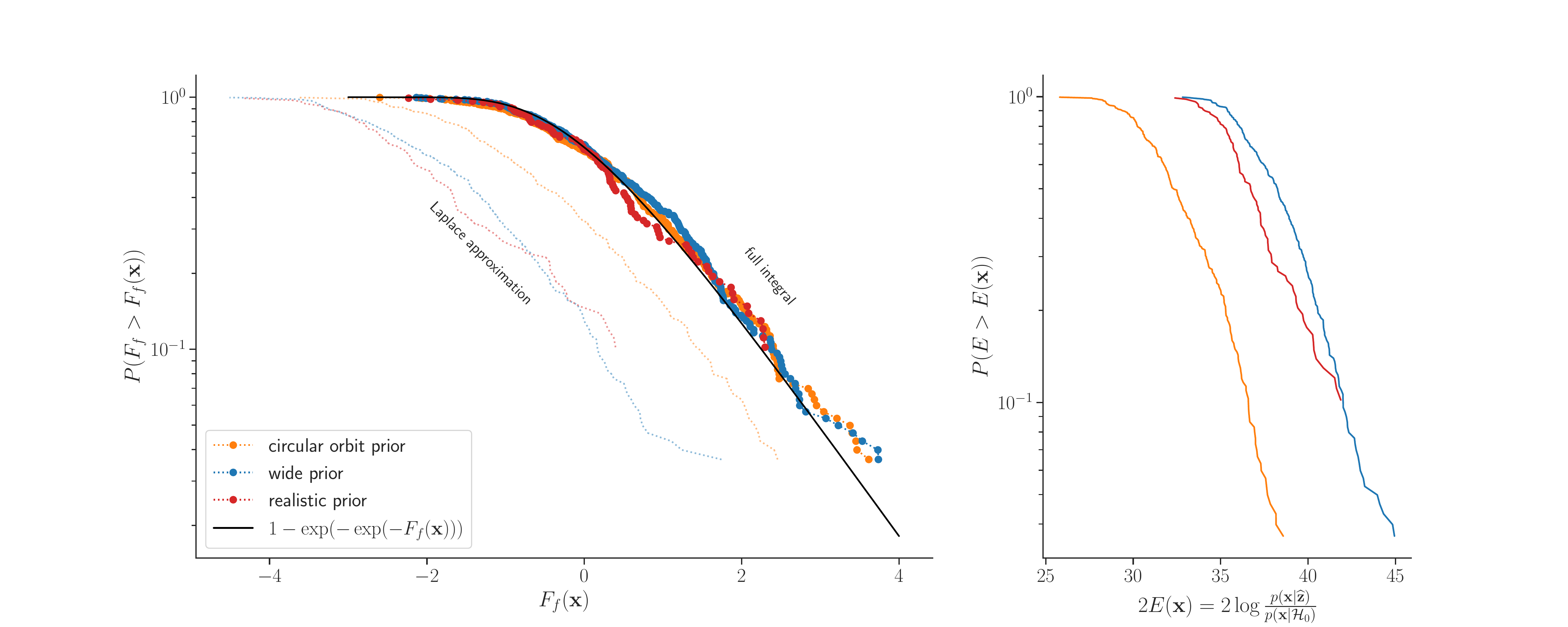}
    \caption{We simulate 300 noise series and search for the planetary signal. In each realization we find a candidate with highest Bayes factor and evaluate it's p-value using Equations \eqref{eq: approx pvalue} and \eqref{eq: sidak}. In the left panel, we then compare it with a fraction of noise realizations where a more significant planet candidate was found. A good agreement between the empirical and analytical p-value is found, regardless of the prior choice. Dotted lines are the corresponding results using the asymptotic limit of Laplace approximation, which is inadequate in this case. In the right panel, we show the likelihood-ratio test statistic distribution for the null hypothesis using the same simulations. Note a large trials factor which causes a noise-only simulation to produce a ${\rm SNR}= \sqrt{2 E} > 6$ in most realizations. There is an order of magnitude difference in the trials factor for the circular prior versus the wide or realistic prior.}
    \label{fig: pvalues}
\end{figure*}

As a more complex non-trivial example of the formalism we developed, we consider exoplanet detections in the transit data, where the 
planet orbiting the star dims the star when it transits 
across its
surface. We have a time series of star's flux measurements $x_i = x(t_i)$. In the absence of a transiting planet, the data is described by a stationary correlated Gaussian noise modeling stellar variability
\footnote{Long-term trends in the Kepler data are removed by the preprocessing module \cite{kepler_process}. Outliers are a source of non-Gaussianities, but they can be efficiently Gaussianized without affecting the planet transits \cite{GMF}. Here we ignore other defects in the data, such as binary stars, sudden-pixel-sensitivity-dropouts, etc.}.
Such data are for example collected by the Kepler Space Telescope \cite{KeplerSpaceTelescope} and Transiting Exoplanet Survey Satellite (TESS) \cite{TESS}.  
\par
One would like to compare the hypothesis $\mathcal{H}_1$ that we have a planet in the data to the null hypothesis that there is only noise. As argued in this paper, we will use the Bayes factor as a test statistic to incorporate informative prior information and non-Gaussian posteriors. The significance of a discovery is then reported as an associated false positive probability
of exceeding the observed value of the Bayes factor. We will first outline the procedure for calculating the Bayes factor given the prior for the transit model parameters. Then, using simulations of the null hypothesis, we will demonstrate that Equation \ref{eq: sidak} gives reliable results for the p-value of the Bayes factor (Subsection \ref{sec: pvalue test}). In Subsection \ref{sec: prior}, we will discuss several prominent prior choices and demonstrate how a realistic prior choice can reduce Type II error at a constant Type I error.

\subsubsection{Bayes factor}\label{sec: Bayes Kepler}

Planet transit model $m(t)$ can be parametrized by $d = 4$ parameters: transit amplitude, period, phase, and transit duration, $\bi{z} = (A, P, \phi,\tau)$. The transit model is of the form
\begin{equation} \label{eq:signal}
    m_i(\bi{z}) = A \sum_{m = 1}^{M} U\bigg( \frac{t_i - (m+\phi) P}{\tau}\bigg).
\end{equation}
It is a periodic train of $M$ transits with period $P$, phase $\phi$, amplitude $A$, and transit duration $\tau$. $U(x)$ is a U-shaped transit template that is nonzero in the region (-1/2, 1/2) and depends on the limb darkening of the stellar surface \cite{simple_limb_darkening}.
\par
The likelihood is Gaussian and stationarity ensures that the Fourier transformation $\mathcal{F}$ diagonalizes the covariance matrix with the power spectrum on the diagonal. The power spectrum $\mathcal{P}(\omega) = \langle \vert \mathcal{F} \{n\}\vert^2 \rangle$ can be learned from the data \cite{FourierGP}.

In the language of Subsection \ref{sec: glik} we introduce
\begin{equation}
    \boldsymbol{x} \xrightarrow[]{} \frac{\mathcal{F} \{ \boldsymbol{x} \} }{\mathcal{P}^{1/2}} \qquad
    \boldsymbol{m} \xrightarrow[]{} \frac{\mathcal{F} \{ \boldsymbol{m} \} }{\mathcal{P}^{1/2}}
\end{equation}
to make the likelihood a standard Gaussian. We also rescale the amplitude $A$ to make the model normalized.

It will be convenient to first optimize over the linear parameter $z_1$ and then over the remaining non-linear parameters. Amplitude's optimal value at fixed $\boldsymbol{z}_{>1}$ is by the Equation \eqref{eq: stationarity}
\begin{equation}
    \widehat{z}_1 (\boldsymbol{z}_{>1}) = \frac{ \int \FT{x}^* \FT{m} d \omega/\mathcal{P}  }{\sqrt{\int \vert \FT{m} \vert^2 d \omega/\mathcal{P} }}.
\end{equation}
where we have restored a model normalization factor. This is the matched filtering expression for the signal-to-noise ratio \cite{GMF} and can be computed efficiently using Fourier transforms \cite{GMF} for all phases $\phi$ on a fine grid at once. Our strategy for finding the MAP solution is therefore to first find a maximum likelihood estimator (MLE) by scanning over the entire parameter space (for details see \cite{GMF}) and then use it as an initial guess in a nonlinear MAP optimization.

We calculate the posterior volume associated with the most promising peaks by marginalizing over the planet's parameters in the vicinity of the peak:
\begin{equation} \label{eq: B Kepler}
V_{\text{post}}^{>1} = \int \frac{p(\boldsymbol{z}_{>1}) \, e^{\widehat{z}_1 (\boldsymbol{z}_{>1})^2 /2}}{p(\zhat) \, e^{\widehat{z}_1^2 / 2}} \, d \boldsymbol{z}_{>1} ,
\end{equation}

One would be tempted to employ the Laplace approximation. Figures \ref{fig: integrand} and \ref{fig: pvalues} show this does not give satisfactory results: we are not in the asymptotic limit. While a frequentist approach using Wilks' theorem would become invalid, here, we can compute the full marginal integral of Bayesian evidence. 
Integral \eqref{eq: B Kepler} is only three-dimensional, so we take the Hessian at the peak to define the Gaussian quadrature integration scheme \cite{gaussquadratureformulas} of degree 7, implemented in \cite{quadpy}, which requires 24 integrand evaluations.

\subsubsection{p-value} \label{sec: pvalue test}

To obtain the p-value, we use Equation \ref{eq: approx pvalue} and extend it to all p-values using \eqref{eq: sidak}.
\par
We test our analytical expression for the p-value of the Bayes factor by simulating the null hypothesis and comparing the computed p-value with the empirically determined value. We evaluate 300 realizations of the null hypothesis. We take a realistic power spectrum extracted from Kepler's data for the star Kepler 90. The power spectrum and example realizations are shown in Figures 2 and 4 of \cite{GMF}.
A realization is a uniformly distributed choice of the phases of the Fourier components and normally distributed amplitudes of the Fourier components with zero mean and variance given by the power spectrum (an example realization is shown in Figure 4 of \cite{GMF}). In each of the resulting time-series, we then determine the Bayes factor of the planet hypothesis against the null hypothesis (Subsection \ref{sec: Bayes Kepler}) and its p-value. 
\par
Analytic and empirical p-value are compared in Figure \ref{fig: pvalues}, achieving an excellent agreement. This shows that our formalism enables the evaluation of p-value beyond the asymptotic limit, thus generalizing Wilks' theorem. 

\subsubsection{The choice of prior} \label{sec: prior}

\begin{figure}
    \centering
    \hspace*{-0.5cm}\includegraphics[scale = 0.27]{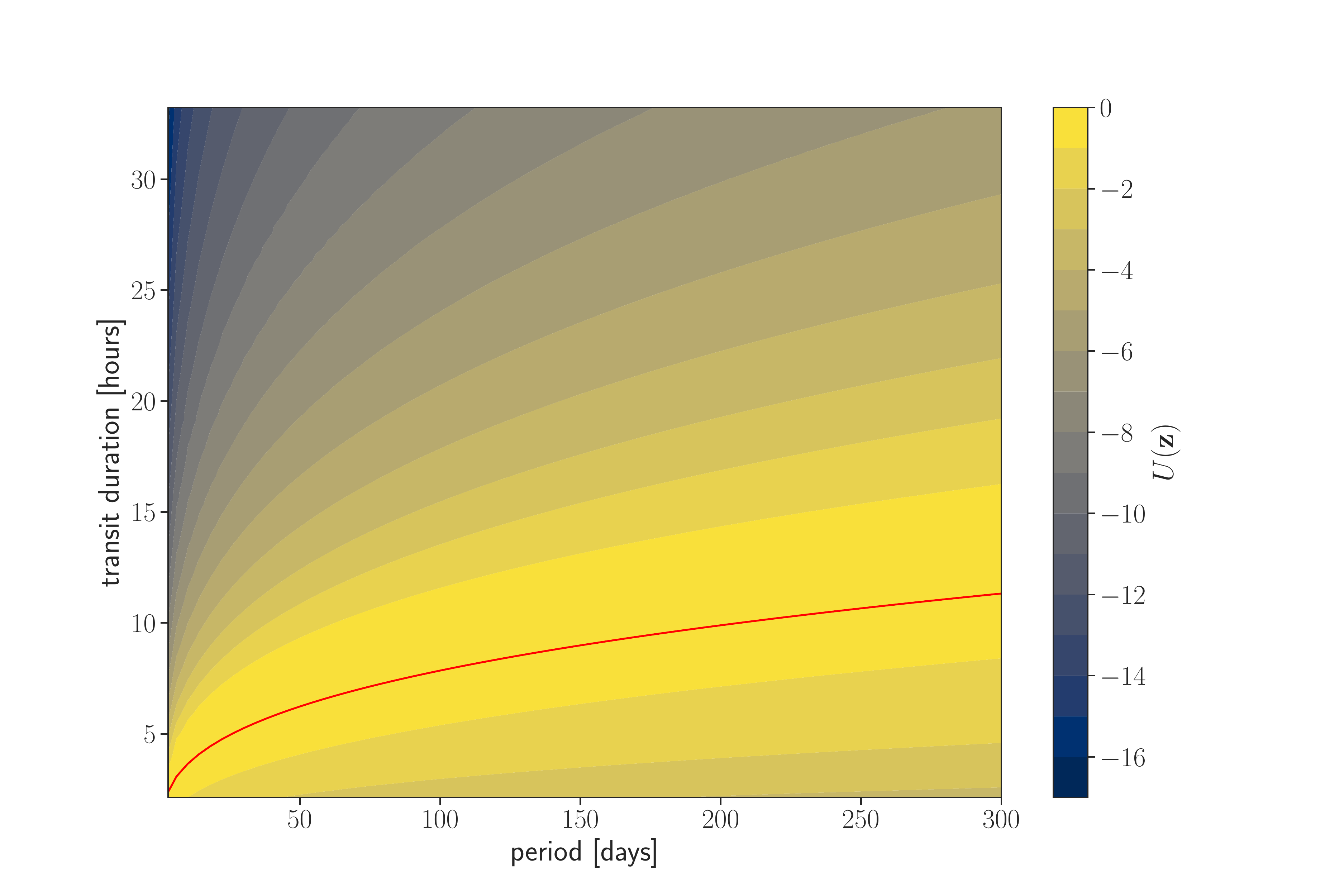}
    \caption{We show the potential $U = \log p(\boldsymbol{z}) / p_{Jeff}(\boldsymbol{z})$ for different choices of the $\tau$ prior. This quantity measures to what extent is the prior informative with the Jeffrey's prior being non-informative. This quantity directly impacts the hypothesis test's outcome, see Section \ref{sec: energy-entropy}. 
    Fixing $\tau$ to the Kepler value is equivalent to a constant potential on the red line and $- \infty$ everywhere else. A broad Jeffrey's prior is equivalent to a constant potential over the entire search domain. The informative prior is shown with the color scale. It peaks close to the Kepler's value, but is broadened to account for variation in planet eccentricity orbit orientation and stellar density uncertainty.}
    \label{fig: potential}
\end{figure}

\begin{figure*}
    \centering
    \includegraphics[scale = 0.35]{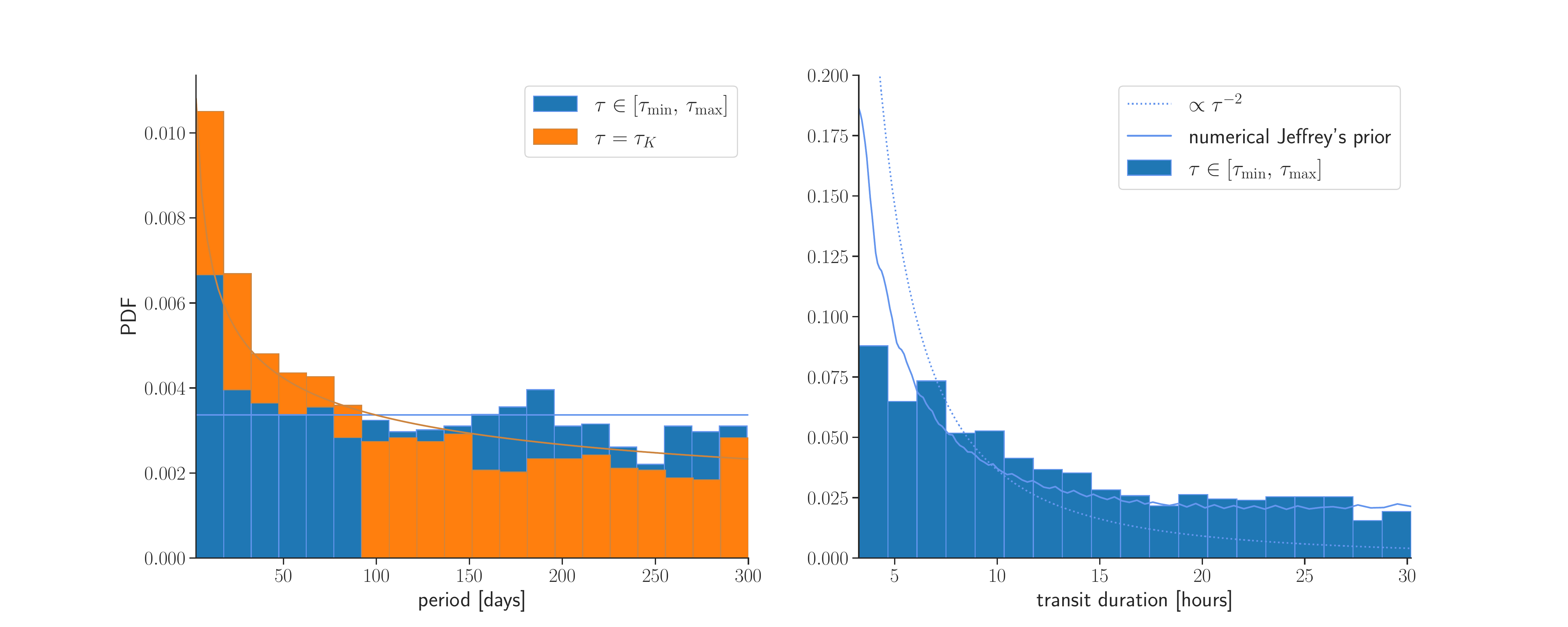}
    \caption{We show marginal probability density distributions over the period and transit duration of the five highest $F_f$ candidates from each of the 300 noise-only simulations. The distributions follow Jeffrey's prior, which is approximated reasonably well with the scaling estimates from Equations \eqref{eq: Jeff1} and \eqref{eq: Jeff2}. This confirms that 
    regions with smaller
    $V_{\rm post}$ have more false positives. 
    The blue plot is showing a situation where the transit duration is a free parameter in the search, yielding $p(P) \sim {\rm const}$ and $p(\tau) \sim \tau^{-2}$. A simple analytical estimate does not give an accurate Jeffrey's prior for the transit duration because of the non-white power spectrum. Therefore, we complement this analysis with a numeric evaluation of the expected posterior volume (solid line) and show that this gives a better match. 
    The orange plot shows a situation where the transit duration is fixed to the Kepler's value. It is well described by the scaling estimate $p(P) \propto P^{-1/3}$.}
    \label{fig: Jeffrey}
\end{figure*}

Here we discuss a prior choice for the period, phase, and transit duration. We will see that using informative prior when available can significantly improve detection efficiency. 
\par
\begin{figure}
    \centering
    \includegraphics[scale = 0.33]{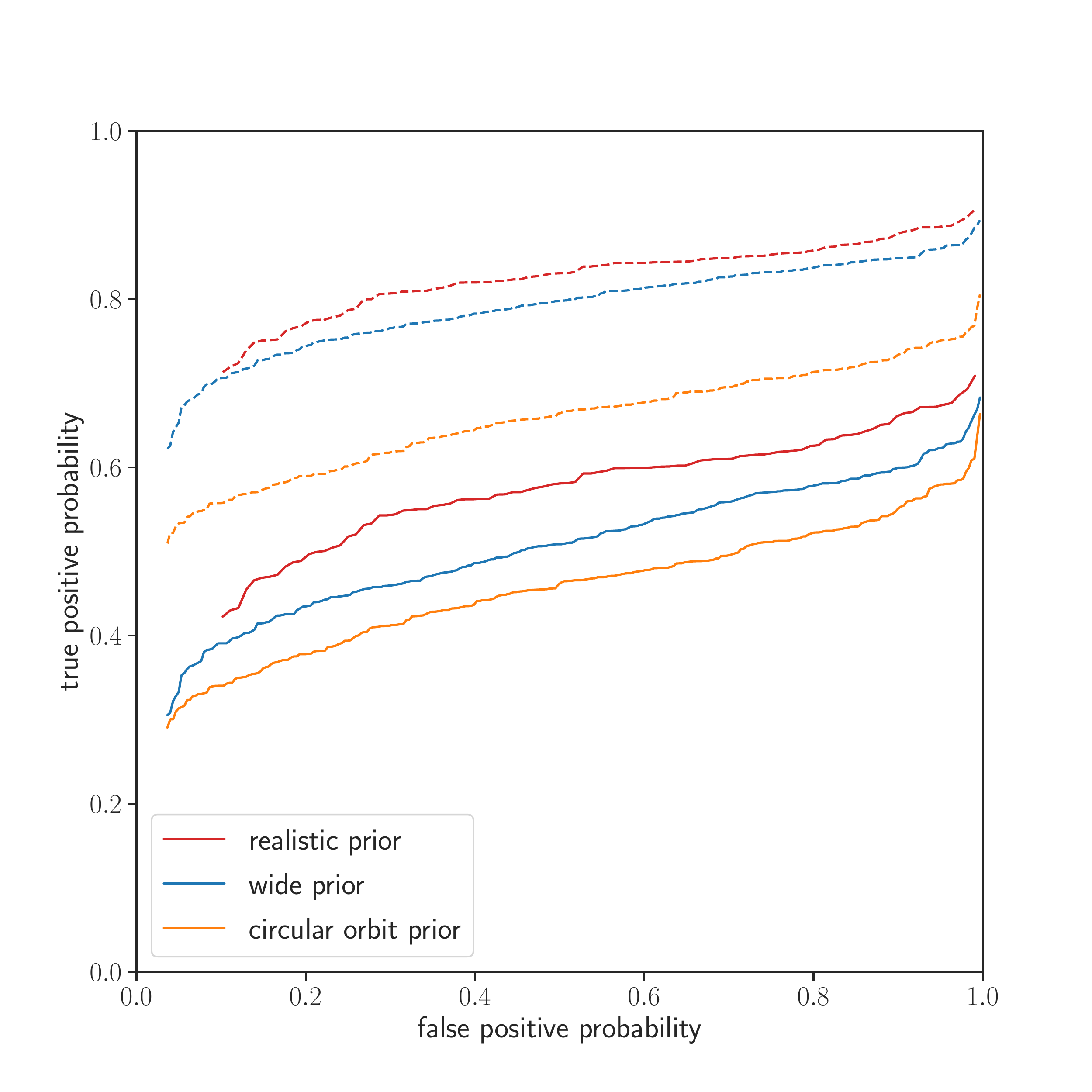}
    \caption{ROC curve (see Subsection \ref{sec: prior}) defined as true positive probability, i.e. power (1-Type II error) versus false positive probability, i.e. Type I error (p-value), for various prior choices. We search for a hypothetical small planet with injected radius $R = 1.45 R_{\bigoplus} $, corresponding to ${\rm SNR}(true) = 6.25$ (solid lines), and $R = 1.8 R_{\bigoplus} $ and ${\rm SNR}(true) = 7$ (dotted lines). Using the realistic prior results in an improved ROC curve (higher true positive probability at a given false positive probability) implying that the Bayes factor with a realistic prior is the test statistic with the highest statistical power. The true positive probability at a fixed p-value is decreased for the circular orbit prior because the prior does not include all the allowed signal template forms. On the other hand, a wide prior includes too many signal template forms that include signal templates that occur rarely or never, leading to a larger multiplicity and a larger false positive rate at a constant true positive probability.}
    \label{fig: Kepler prior downgrades}
\end{figure}

We scan over the period range from 3 to 300 days and at each period over all phases from 0 to 1. As we argued in this paper, 
when we do not know the prior, 
it is simplest to adopt Jeffrey's choice. We will assume this for period and phase parameters. 
Note that for the phase, Jeffrey's prior is flat 
from 0 to 1, which we can also view
as a known (informative) prior, since the phase
cannot affect the exoplanet detectability. 
There is a natural value for the transit duration parameter $\tau$, given the planet's period and assuming the circular, non-inclined orbit. 
We illustrate the impact of the prior choice by considering three scenarios:\newline
    1) A fixed value of $\tau$, defined by assuming a circular, perfectly aligned orbit. The transit duration is fixed by the Kepler's third law: $\tau_K(P) = q\,  P^{1/3}$, with the proportionality constant given by the star's density, $q = (3 P / (\pi^2 G \rho_*))^{1/3}$. This choice can be too restrictive for non-circular or inclined orbits and can penalize real planets, as shown in Figure \ref{fig: Kepler prior downgrades}. Under this prior choice, the Bayes factor is proportional to the likelihood ratio.\newline
    2) Jeffrey's prior on a broad domain $\tau \in [\tau_K(P_{min}), \, 2 \, \tau_K(P_{max})]$. This choice may include physically implausible transit times leading to large multiplicity 
    penalty.
    \newline
    3) realistic prior distribution, taking into account inclined and eccentric orbits. Orbits are assumed to be isotropic, with eccentricities drawn from a beta function with parameters that match the observed planets in the Kepler's data \cite{EccentricityPriorKipping}. In addition, the star's density is measured with an uncertainty on the order of 15 \%, causing uncertainty in $\tau_K$. The transit duration prior is obtained by marginalizing over the orbit inclination, eccentricity and star's parameters. The distribution is a broadened version of the delta function at $\tau_K$.
    
All three choices are visualized in Figure \ref{fig: potential}.
In appendix \ref{appendix: Jeffrey} we derive the approximate Jeffrey's prior by deriving the scaling of the Fisher information matrix. In the first two scenarios, we respectively get $p \propto \tau^{-2}$ and $p \propto P^{-1/3}$. One can take the template and calculate the Fisher information matrix numerically for more accurate results. In Figure \ref{fig: Jeffrey} we show that peaks of the null hypothesis are distributed by Jeffrey's prior, in agreement with our predictions.
\par
Taking $\tau$ as a completely unconstrained parameter will in general lead to detecting planets that are not physically plausible and will therefore increase the false positive rate and force us to reject more real planets than necessary. In the present example shown in 
figure \ref{fig: pvalues} this effect is not very large if the prior is chosen such that it still introduces a reasonable cutoff on the transit duration. But choosing such a cutoff is
a choice that must be based on physical arguments: the prior plays an essential role. 

A prior that is too narrow (case 1) is also suboptimal because it will reject some real planets. Using a template with duration $\tau_K$ when in fact a planet has transit duration $\tau \neq \tau_K$ will reduce planet's $SNR$ by 
\begin{equation} \label{eq: SNR reduction}
    \frac{{\rm SNR (detected)}}{{\rm SNR(true)}} = \frac{\int \FT{m_{\tau}}^* \FT{m_{\tau_K}} d \omega/\mathcal{P}}{\int \FT{m_{\tau_K}}^* \FT{m_{\tau_K}} d \omega/\mathcal{P} },
\end{equation}
where this is an expected reduction over the noise realizations.
\par
In Figure \ref{fig: Kepler prior downgrades} we show the ROC curves (Receiver Operated Characteristic) for all three prior choices, that is, a true positive rate as a function of the false positive rate, both parametrized by the detection threshold. False positive probabilities are taken from Figure \ref{fig: pvalues}. True positive probability is a probability that the signal with some ${\rm SNR(true)}$ is detected above the threshold. The detected ${\rm SNR}$ can be 
approximated as a Gaussian distributed variable with unit variance. Its mean is $\mu = {\rm SNR(true)}$ with the realistic and wide priors. It is additionally reduced with the circular orbit prior because the search template differs from the true template (Equation \eqref{eq: SNR reduction}). Using a realistic prior improves the true detection probability at a fixed false positive probability relative to a prior that is too narrow (circular orbit prior) because it improves the fit. It also improves ROC relative to a prior 
that is too broad because it does not include the
templates that rarely happen in the data and would
lead to a larger multiplicity and hence a larger 
false positive rate. This figure thus demonstrates that hypothesis testing with a realistic prior Bayes factor gives optimal ROC (power versus p-value).

\section{Discussion}

This paper compares frequentist and Bayesian 
significance testing between hypotheses of different dimensionality, where the null hypothesis is a well-defined accepted model for the reality of the data, while the alternative hypothesis tries to 
replace the null hypothesis. Both Bayesian and frequentist methodologies have advantages and disadvantages in the setting where the 
alternative hypothesis has not been observed with sufficient frequency to develop a reliable prior. 
We argue that for optimal significance testing, the Bayes factor
between the two hypotheses should be used as the
test statistic with the highest power. 
However, the Bayes factor should not be used to quantify the test significance when the prior of the alternative hypothesis is poorly 
known. Instead, the frequentist false positive rate of null 
hypothesis (Type I error or p-value) can be used, which 
only depends on the properties of the null 
hypothesis, which is assumed to be well 
understood. The sensitivity of the Bayes factor to 
the choice of prior is known in the context of Lindley's paradox \cite{Lindley57}. While there is no actual 
paradox, it highlights the dependence of the 
Bayes factor on the choice of prior, which is 
undesirable since we often do not know it. 
Our solution to this paradox is to relate the 
Bayes factor to the p-value, which is independent 
of the alternative hypothesis, as it only tests
the distribution of the null hypothesis. In this 
way one can use the p-value for hypothesis testing even when 
using Bayesian methods, by using the Bayes factor  
as a test statistic. 

While it is common to use the 
likelihood ratio as a test statistic, this is 
not prior independent, but corresponds to the Jeffrey's prior. If some prior information is available, as in our exoplanet example of transit duration being determined by the 
period $P$ via the Kepler law, 
one should use it to reduce the Type II error. 
We note that 
Jeffrey's prior can be unreasonable even in 
unknown prior situations: 
if, in a given experiment, the posterior volume is 
strongly varying across the parameter space, the Jeffrey's prior is very experiment specific. A prior that is smooth across the parameter space is undoubtedly a better prior, even if we do not know what the specific form should be. Nevertheless, when Jeffrey's prior is reasonable, it simplifies the analytic calculation of the p-value. 

We show that both the p-value and the Bayes factor can be expressed as energy versus entropy competition. We define energy as the logarithm of the likelihood ratio and Bayesian entropy as the logarithm of the number of posterior volumes that fit in the prior volume. The constant energy Bayes factor is analogous to the thermodynamical free energy. Conversely, the p-value in the asymptotic regime corresponds to the canonical partition function, which is a Boltzmann factor weighted sum over the posterior states with the test statistic above the observed one. In the low p-value regime, only the states close to the observed test statistic contribute to the sum. Therefore, the constant energy Bayes factor and the asymptotic p-value are related. This also happens in physics, where the statistical and thermodynamical definitions of the free energy coincide in the thermodynamical limit.
As an example, we show that the p-value of the standard $\chi^2$ distribution of $d$ degrees of freedom can be interpreted as energy versus entropy competition, with the latter defined as the logarithm of the number of states on the constant energy shell. Entropy grows as the log of the area of a sphere in $d$ dimensions with a radius proportional to the square root of energy. 

The formalism developed here extends the Wilks' theorem \cite{wilksTheorem} in several ways. 
First, the connection to the Bayes factor allows us to define the posterior volume beyond the Gaussian approximation inherent in the asymptotic limit assumed for the Wilks' theorem. 
Second, the Wilks' theorem assumes the parameter values are inside the boundaries. We show that a generalization counting the states as a function of energy provides a proper generalization 
that gives better results. Third, Wilks' theorem does not account for 
the multiplicity of the Look Elsewhere Effect and thus fails for coordinate
parameters, which do not change the expected energy. Our method correctly handles these situations. 

As an example application, we apply the formalism to the exoplanet transit search in the stellar variability polluted data. We search for 
exoplanets by scanning over the period, phase, and 
transit duration and we show that the multiplicity from the Look Elsewhere effect is of order $10^7$. We find that the Laplace approximation for uncertainty volume $V_{\rm post}$ is inaccurate, while the numerical integration of the Bayesian evidence gives very accurate results when compared to simulations. We emphasize the role of informative priors such as planet transit time prior, which reduce the Type II error, leading to a higher fraction of true planets discovered at a fixed Type I error (p-value) threshold. The method
enables fast evaluation of the false positive 
rate for every exoplanet candidate without 
running expensive simulations. 

There are other practical implications that follow from our analysis. For example, the multiplicity depends not only on the prior range but also on the posterior error on the scanning parameters. If this error is small in one part of the parameter space but large in others, then the likelihood ratio test statistic leads to a large multiplicity that will increase the p-value for all events. We show that analytic predictions reproduce the distribution of false positives as a function of period and transit duration. 
An informed choice of the prior 
guided by what we know about the problem and 
what our goals are may 
change this balance and reduce the multiplicity 
penalty, thus reducing the Type II error: Bayes factor can be a better test statistic for Type II error than the likelihood ratio. One is 
of course not allowed to pick and choose the prior a 
posteriori: we must choose it prior to the data analysis. 

In many situations, it is possible to analytically obtain
the false positive rate as a function of 
Bayes factor test statistic from Equation 
\eqref{eq: approx pvalue}, which gives a p-value 
estimate that is more reliable for 
hypothesis testing than the 
corresponding Bayes factor in situations of 
a new discovery where the prior is not yet known.  
As a general 
recommendation, we thus 
advocate that Bayesian analyses 
report the frequentist p-value using the Bayes factor 
test statistic against the null hypothesis 
as a way to quantify the 
significance of a new discovery, 
and that frequentist analyses 
use the Bayes factor as the optimal test statistic for hypothesis 
testing while using frequentist methods to quantify its significance.

\vspace{6pt} 




\acknowledgments{We thank Adrian Bayer and Natan Dominko-Kobilica for many valuable discussions and comments.
This material is based upon work supported in part by the Heising-Simons Foundation grant 2021-3282 and by the U.S. Department of Energy, Office of Science, Office of Advanced Scientific Computing Research under Contract No. DE-AC02-05CH11231 at Lawrence Berkeley National Laboratory to enable research for Data-intensive Machine Learning and Analysis.
}

\appendix
\section{Jeffrey's prior in the exoplanet example} \label{appendix: Jeffrey}
Here, we derive the approximate Jeffrey's prior in the exoplanet example by analyzing the Fisher information matrix \eqref{eq: Fisher def}:
\begin{equation} \label{eq:FisherIntegral1}
 I_{ij}({\bf z}) =  Re \int \frac{\partial \FT{m}}{\partial z_i}\frac{\partial \FT{m}^*}{\partial z_j} \frac{d\omega}{P(\omega)} ,
 \end{equation}
Let us first determine the scaling of the $I_{\phi \phi}$. Fourier transforming the signal from Eq.~\ref{eq:signal}, differentiating it with respect to $\phi$, inserting into equation  \ref{eq:FisherIntegral1} and expressing the amplitude $A = z_1$ with $E$ we obtain
\begin{equation}
    I_{\phi \phi} = 2 E \frac{\int \frac{\omega^2 P^2 \vert\FT{m} \vert^2d \omega}{P(\omega)}}{\int \frac{\vert\FT{m} \vert^2d \omega}{P(\omega)}} \propto \frac{P^2}{\tau^2} \, ,
\end{equation}
where in the second step we change the variable of integration to the dimensionless quantity $\omega \tau$ and neglect a residual mild dependence of the dimensionless integral on $P$ and $\tau$.
\par
In a scenario where $\tau$ is an independent quantity we similarly get $I_{PP} \propto P^{-2} \tau^{-1}$ and $I_{\tau \tau} \propto \tau^{-1}$, therefore the Jeffrey's prior is:
\begin{equation} \label{eq: Jeff1}
    p({\bf z}) \propto \det I ^{1/2} \propto \tau^{-2} \, .
\end{equation}
In the other scenario, where $\tau$ is fixed by the period we have $I_{PP} \propto P^{-2}$ and Jeffrey's prior is:
\begin{equation}\label{eq: Jeff2}
    p({\bf z}) \propto \tau_{K}^{-1} \propto P^{-1/3} \, .
\end{equation}
An estimate of the Jeffrey's prior is obtained by normalizing in the given range of the parameters.

\bibliographystyle{unsrt}
\bibliography{citations}

\end{document}